\documentclass[journal]{IEEEtran}
\usepackage{algorithm}
\usepackage{url}
\usepackage{amssymb}
\usepackage{amsmath}
\usepackage{epsfig}
\usepackage{epstopdf}
\usepackage{subfigure}
\usepackage{balance}
\usepackage{multirow}
\usepackage{booktabs}
\usepackage{color, colortbl}
\usepackage{tabularx}
\usepackage{algcompatible}
\usepackage{soul}
\usepackage{xcolor}
\usepackage{amsmath}
\usepackage{amssymb}
\usepackage{mathrsfs}

\DeclareMathOperator*{\argmax}{arg\,max}

\ifCLASSINFOpdf
\else
\fi
\begin{document}
%

\title{Distributed Energy Trading and Scheduling among Microgrids via Multiagent Reinforcement Learning}

%
%

\author{Guanyu~Gao,
        Yonggang~Wen,~\IEEEmembership{Fellow,~IEEE},
        Xiaohu~Wu~and~Ran~Wang 
\thanks{Guanyu Gao, Yonggang Wen and Xiaohu Wu are with the School of Computer Science and Engineering, Nanyang Technological University, Singapore. 
Ran Wang is with the School of Computer Science and Engineering, Nanjing University of Aeronautics and Astronautics, Nanjing, China.
Email: \{ggao001, ygwen, xiaohu.wu\}@ntu.edu.sg, wangran@nuaa.edu.cn.
}
}
%
%
%

\markboth{Journal of \LaTeX\ Class Files,~Vol.~14, No.~8, August~2015}%
{Shell \MakeLowercase{\textit{et al.}}: Bare Demo of IEEEtran.cls for IEEE Journals}
%



\maketitle

\begin{abstract}
The development of renewable energy generation empowers microgrids to generate electricity to supply itself and to trade the surplus on energy markets.
To minimize the overall cost, a microgrid must determine how to schedule its energy resources and electrical loads and how to trade with others.
The control decisions are influenced by various factors, such as energy storage, renewable energy yield, electrical load, and competition from other microgrids.     
Making the optimal control decision is challenging,
due to the complexity of the interconnected microgrids, the uncertainty of renewable energy generation and consumption, and the interplay among microgrids.
The previous works mainly adopted the modeling-based approaches for deriving the control decision,
yet they relied on the precise information of future system dynamics, 
which can be hard to obtain in a complex environment.
This work provides a new perspective of obtaining the optimal control policy for distributed energy trading and scheduling by directly interacting with the environment, 
and proposes a multiagent deep reinforcement learning approach for learning the optimal control policy. 
Each microgrid is modeled as an agent, and different agents learn collaboratively for maximizing their rewards.
The agent of each microgrid can make the local scheduling decision without knowing others' information, which can well maintain the autonomy of each microgrid.
We evaluate the performances of our proposed method using real-world datasets.
The experimental results show that our method can significantly reduce the cost of the microgrids compared with the baseline methods.

\end{abstract}


%
\IEEEpeerreviewmaketitle

\section{Introduction}
Distributed energy generation enables the microgrids to generate electricity from renewable sources, such as solar radiation, winds, etc.
This can alleviate the environmental pollution caused by fossil fuel-based generation and reduce the energy loss caused by long-distance transmission \cite{lasseter2011smart}.
The microgrids can save the cost by utilizing renewable energy 
and obtain revenue by selling the surplus in energy markets.
Therefore, it has been witnessed that more and more renewable energy generators are deployed in the microgrids.

This paradigm has significantly changed the ways of the trading and scheduling of energy markets \cite{mengelkamp2018designing}.
Each microgrid has become an autonomous entity, which can determine the trading price and quantity with other microgrids based on its energy demands and yields.
On the other hand, it also requires each microgrid to make appropriate control decisions on how to schedule its electrical load and energy storage.
For example, a microgrid may choose to charge the battery when the market price is low,
and discharge the battery for utilization when the energy market price is high.
Therefore, the microgrids must have some control mechanism for making optimal trading and scheduling decision to maximize their interests.

Making the trading and scheduling decisions for a microgrid is challenging from the following aspects.
First, the trading decision of a microgrid is not only affected by itself but also by other microgrids.
In a bidding market, if a microgrid's selling price is too high or buying price is too low, no deal can be made, which will, therefore, affect the cost and the revenue of the microgrid. 
Second, the trading and the scheduling decisions of a microgrid are coupled.
Making the decisions for trading and scheduling should consider every aspect of the whole system, including the electrical loads, energy storage and yield, and market price.
Third, the intermittency of renewable energy sources (e.g., solar irradiation and wind speed) increases the uncertainties of energy generation,
which makes the control problem more complex.

Some existing works separately studied the energy trading and scheduling problem for microgrids.
Several previous works (e.g., \cite{gregoratti2014distributed,matamoros2012microgrids,wang2014game,lee2015distributed,wang2016reinforcement}) adopted different approaches to design the trading strategies.
However, they did not take into account the scheduling of the local energy storage and the electrical load in each microgrid.
Some other works (e.g., \cite{fathi2013adaptive,huang2014adaptive,fathi2013statistical}) considered the energy storage and load scheduling problems for microgrids.
Yet, they considered the problem either from the perspective of an individual microgrid or from the perspective of a centralized controller, which assumes that all microgrids work as an entire entity without autonomy. 
Meanwhile, these works did not consider energy trading among the microgrids.
These two lines of works either only considered the energy scheduling problem for microgrids, or only considered the trading problems for microgrids.
Thus, they lack a comprehensive and all-sided analysis for the trading and scheduling problem for microgrids.

The works in \cite{wang2015bargaining,paudel2018peer,kim2019direct,li2018distributed} jointly considered the energy trading and scheduling problem among microgrids for maximizing the interest of each microgrid.
These works mainly adopted the game-theoretic models for modeling the trading and scheduling for microgrids \cite{tushar2020peer}.
The models rely on the precise information of future energy yields and electrical loads as inputs for deriving the optimal scheduling solution.
However, different microgrid environments may be full of various uncertainties regarding renewable energy generation and energy consumption, which affects the performances of the scheduling.   
For instance, a microgrid may plan to sell a certain quantity of renewable energy, however, the contract may not be fulfilled because of bad weather conditions.
The complexity of the microgrid system, the uncertainty of the environments, and the interplay among different microgrids posit many challenges on precisely modeling the problem and deriving the optimal control decision.

The multiagent deep reinforcement learning approach provides a new perspective for learning the optimal control policy by directly interacting with the environment.
In this paper, we propose a multiagent deep reinforcement learning approach \cite{sutton2018reinforcement} for distributed energy trading and scheduling for microgrids.
Each microgrid is modeled as an agent, and different agents may collaborate or compete with each other for maximizing its reward.
We adopt the Multi-Agent Deep Deterministic Policy Gradient (MADDPG) approach \cite{lowe2017multi}  to design the algorithm for learning the optimal control policy for each microgrid.
After training, the agent of each microgrid can independently make the control decision without requiring the information from other microgrids.
The experimental results from extensive performance evaluations show that our method can significantly reduce the cost of each microgrid.

The main contributions are summarized as follows.
\begin{itemize}
\item
Propose a multiagent deep reinforcement learning approach for modeling the distributed energy trading and scheduling problem for microgrids and learning the optimal control policy for each microgrid. 

\item
Design a bidding-based incentive mechanism for energy trading among microgrids. 
The mechanism is suitable for multiagent reinforcement learning based energy trading.

\item
Design the network models for the actor and the critic based on the characteristics of the control actions  and the role of energy trading of each microgrid.

\item
Implement an energy trading and scheduling simulation environment and evaluate the performances of the proposed method using real-world datasets.
\end{itemize}

The rest of this paper is organized as follows. 
Section \ref{sec:related-work} introduces the related works,
Section \ref{sec:system-overview} presents the overview of the energy trading and scheduling system,
Section \ref{sec:system-model} gives the system models and problem formulation,
Section \ref{sec:algorihtm} designs the neural networks and the learning algorithm,
Section \ref{sec:performance} evaluates performances,
and Section \ref{sec:conclusion} concludes the paper.

\section{Related Work} \label{sec:related-work}
In this section, we review the existing works on energy trading and scheduling among microgrids.

One line of research adopted different methods to model the energy trading for microgrids.
Gregoratti \emph{et al.} \cite{gregoratti2014distributed} and Matamoros \emph{et al.} \cite{matamoros2012microgrids} considered the problem of energy trading among islanded microgrids and adopted the distributed convex optimization framework for minimizing cost.
Wang \emph{et al.} \cite{wang2014game} and Lee \emph{et al.} \cite{lee2015distributed} modeled the energy trading among microgrids using a game-theoretic model and formulated the energy trading as a non-cooperative game. 
Wang \emph{et al.} \cite{wang2016reinforcement} proposed an energy trading framework based on the repeated game for maximizing the revenue of each microgrid and adopted reinforcement learning to learn the policy.
These works did not consider the local electrical loads in a microgrid when studying the energy trading problem, yet the scheduling of local electrical loads also affects the trading decision.

Another line of research studied the energy scheduling problem for microgrids. 
Huang \emph{et al.} \cite{huang2014adaptive} developed a control framework for microgrid energy management by categorizing energy usage.
Fathi \emph{et al.} \cite{fathi2013adaptive} proposed an adaptive energy consumption scheduling method for microgrids under demand uncertainty to achieve low power generation cost and low peak-to-average ration.
Wang \emph{et al.} \cite{wang2016cooperative} proposed a framework for planning the renewable energy generation in microgrids. 
Fathi \emph{et al.} \cite{fathi2013statistical} considered the power dispatching in microgrids for minimizing the overall power generation and transmission cost.
Weng \emph{et al.} \cite{weng2018distributed} studied the distributed cooperative control for frequency and voltage stability and power sharing in microgrids.
Ahn \emph{et al.} \cite{ahn2017distributed} proposed the distributed coordination laws for energy generation and distribution in energy network.
These works only considered the energy scheduling problem, and they did not consider the influence of energy trading among microgrids.

The works in \cite{wang2015bargaining,paudel2018peer,kim2019direct,li2018distributed} 
jointly considered the energy trading and scheduling for microgrids.
Wang \emph{et al.} \cite{wang2015bargaining,wang2016incentivizing} proposed a Nash bargaining-based energy trading method.
The proposed method considers each microgrid as an autonomous entity, which aims to minimize its cost through energy bargaining. 
Paudel \emph{et al.} \cite{paudel2018peer} proposed a game-theoretic model for P2P energy trading among the prosumers in a community.
Kim \emph{et al.} \cite{kim2019direct} considered direct energy trading among microgrids under the network's operational constraints.
Li \emph{et al.} \cite{li2018distributed} proposed a bilateral energy trading framework to increase the economics benefits of each individual.
Different from these works, we propose a multiagent deep reinforcement learning approach which can directly learn the optimal control policy for each microgrid by interacting with the environment.

\begin{figure}
\begin{center}
\epsfig{file=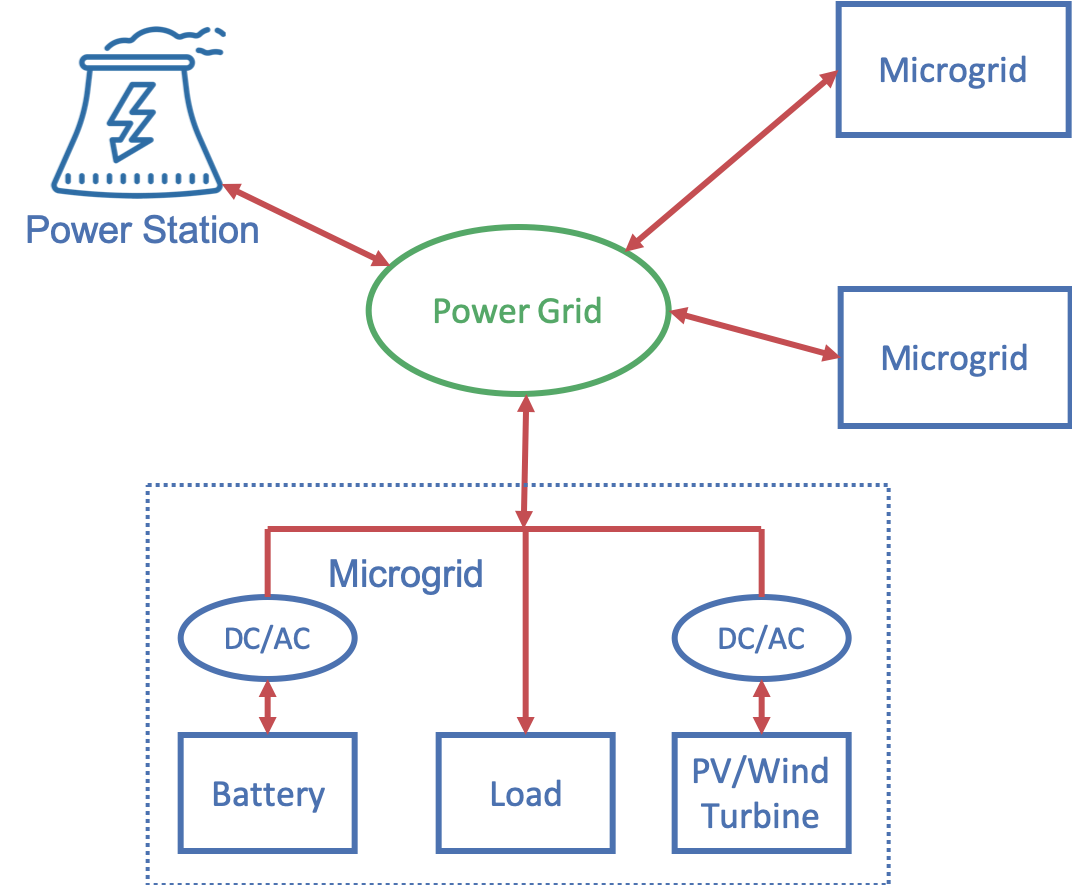, width=0.8\columnwidth}
\end{center}
\caption{The reference system of interconnected microgrids. Several microgrids are interconnected with each other for energy trading and scheduling.} 
\label{fig-system-design}
\end{figure}

\section{System Overview} \label{sec:system-overview}
In this section, we introduce the system overview for distributed energy trading and scheduling among microgrids.

\subsection{Reference System}
We illustrate the reference system of interconnected microgrids for energy trading and scheduling in Fig. \ref{fig-system-design}.
Several microgrids are interconnected with each other through power delivery system.
Each microgrid consists of the following components: electrical loads, batteries, and renewable energy generators.
The electrical loads are equipments which consume electricity, 
the batteries are utilized to store surplus electricity.
The renewable energy generator can generate electricity from renewable energy sources, such as solar energy or wind energy.
A microgrid can purchase electricity from the main grid on the wholesale market or from other microgrids on the hour-ahead energy market.
It can also its sell its surplus energy to other microgrids on the hour-ahead energy market.

\subsection{Energy Trading Flow}
\begin{figure}
\begin{center}
\epsfig{file=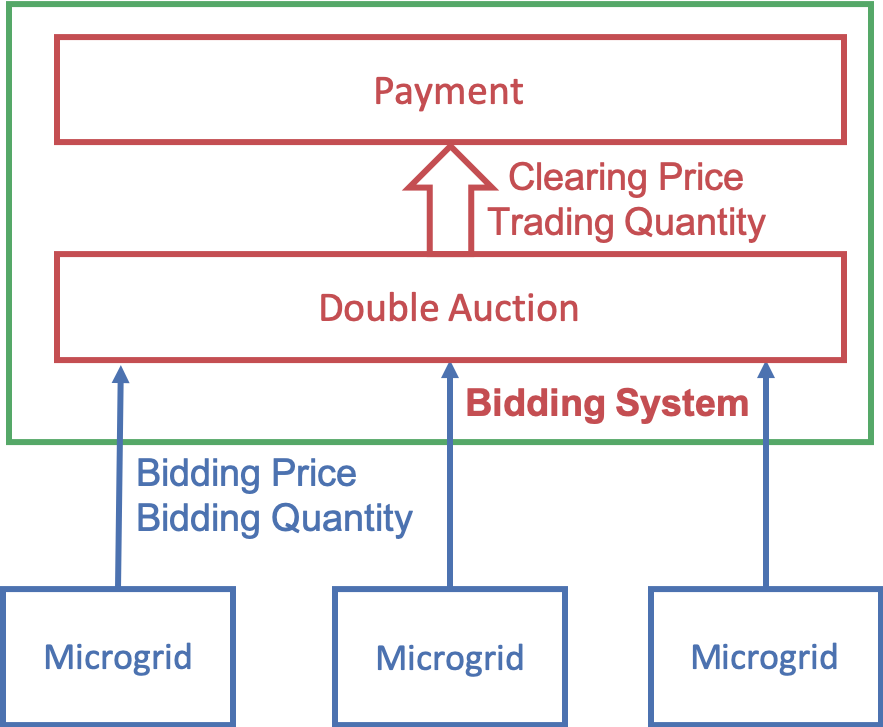, width=0.7\columnwidth}
\end{center}
\caption{The trading flow of the microgrids. Each microgrid submits the bidding price and quantity to the bidding system, and the clearing price and trading quantity will be determined based on the double-auction algorithm.} 
\label{fig-workflow}
\end{figure}

We consider two types of energy markets, namely, the wholesale market and the hour-ahead bidding market.
We illustrate the trading flows for the microgrids on the hour-ahead market in Fig. \ref{fig-workflow}.
At the beginning of each time slot, a microgrid needs to determine how to schedule the local electrical loads, energy storage, and the bidding price and quantity on the hour-ahead market.
The electricity from the power station is sold on the wholesale market and the price is known in advanced.
A microgrid can trade with other microgrids on the hour-ahead bidding market.
Each microgrid determines its bidding (buying or selling) price and quantity. 
The bidding price and quantity of each microgrid will be submitted to the bidding system without disclosing to others.
When the bidding period elapses, the bidding system will calculate the final market clearing price and the trading quantity of each microgrid based on the adopted double-auction trading algorithm.
The payment for each period will be made based on the real electricity usage of each microgrid.

\begin{table}
\centering
\caption{Main Notations} \label{table:notations}
\begin{tabular}{p{0.8cm} p{6.5cm}} \hline \hline
$N$                     & the number of microgrids         \\
$M_i$                   & the i-th microgrid, $i=1,2,...,N$ \\ 
$t$                     & discrete time slot, $t=0,1,2,...$ \\
\hline
$g_i(t)$                & generation power of microgrid $i$ during time slot $t$\\
$r_i(t)$                & solar radiation in microgrid $i$ during time slot $t$ \\
$\psi_i$                & the overall area of the solar panels in microgrid $i$  \\
$e_i$                   & conversion efficiency of the solar panels in microgrid $i$  \\
$B_i$                   & maximum battery storage capacity of microgrid $i$ \\
$b_i(t)$                & battery level of microgrid $i$ at beginning of time slot $t$\\
$l_i(t)$                & electrical load of microgrid $i$ during time slot $t$\\
\hline
$\beta_i^c, \beta_i^d$  & battery charging/discharging efficiency of microgrid $i$\\
$c_i(t)$                & battery charging amount of $M_i$ during time slot $t$ \\
$d_i(t)$                & battery discharging amount of $M_i$ during time slot $t$ \\
\hline
$\Lambda_i(t)$          & a buyer's bid price at time slot $t$ \\
$\Gamma_i(t)$           & a seller's sell price at time slot $t$ \\
$p^*(t)$                & the market clearing price at time slot $t$ \\
\hline 
$\chi_i(t)$             & the bid quantity of microgrid $i$ at time slot $t$ \\
$\rho_i(t)$             & the payment of $M_i$ to the hour-ahead market \\
\hline
$\sigma_i(t)$           & revenue for microgrid $i$ during time slot $t$ \\
$v_i(t)$                & the supplied energy from microgrid $i$ during time slot $t$ \\
\hline
$q_i(t)$                & penalty charged to microgrid $i$ at time slot $t$ \\
$u_i(t)$                & selling quantity of microgrid $i$ during time slot $t$ \\
$f$                     & the weight of penalty for energy under-supply to the grid \\
\hline
$\varpi_i(t)$            & consumed energy from wholesale market by microgrid $i$  \\ 
$\xi_i(t)$               & energy cost of microgrid $i$ for wholesale market  \\  
$p_w$                   & the electricity price of the wholesale market \\
\hline
$s_i(t)$                & the state of microgrid $i$ at time slot $t$ \\
$a_i(t)$                & the control action of microgrid $i$ for time slot $t$ \\
$r_i(t)$                & the reward for microgrid $i$ during time slot $t$ \\
$\phi_i(t)$             & submitted selling quantity of microgrid $i$ at time slot $t$ \\
$\lambda_i(t)$          & submitted buying quantity of microgrid $i$ at time slot $t$ \\
$\pi^{*}_i$             & the optimal control policy of agent $i$ \\
$\gamma_i$              & the discount factor for the reward of agent $i$ \\ 
\hline
$\mu_i(\cdot)$          & the actor network of agent $i$ \\
$Q_i(\cdot)$            & the critic network of agent $i$ \\  
$\mathcal{N}$           & the exploration noise for training \\  
$y^j_i$                 & the target Q-value of transition $j$ for agent $i$ \\     
\hline \hline \end{tabular}
\end{table}

\section{System Model and Problem Formulation} \label{sec:system-model}
In this section, we introduce the system models and problem formulation.
The main notations are summarized in Table \ref{table:notations}.

\subsection{Microgrid Components}
We consider $N$ interconnected microgrids.
The $i$-th microgrid is denoted as $M_i$, where $i = 1,2,...,N$.
We adopt a discrete time system in which the time is denoted as $t=0,1,2,...$.
The duration of a time slot is one hour.
Each microgrid consists of the following components.

\paragraph{Renewable Energy Generation}
Renewable energy can be generated in each microgrid from natural energy sources.
In this work, we consider the renewable energy generation with Solar Photovoltaics (PV).
For simplicity, we assume that the power of the generated electricity in a microgrid is mainly determined by the solar irradiation, the overall area and the conversion efficiency of the solar panels in a microgrid, 
which follows the following equation,  
\begin{equation}
g_i(t) = \psi_i e_i r_i(t),
\end{equation}
where $g_i(t)$ is the average power of electricity generation in microgrid $i$ during time slot $t$, 
$r_i(t)$ is the average solar radiation in microgrid $i$ during time slot $t$,
$\psi_i$ is the overall area of the solar panels in microgrid $i$, 
and $e_i$ is the conversion efficiency of the solar panels in microgrid $i$. 

\paragraph{Battery}
We denote the maximum battery capacity of microgrid $i$ as $B_i$.
The battery level of microgrid $i$ at the beginning of time slot $t$ is denoted as $b_i(t)$.
The battery level changes due to charging and discharging.

\paragraph{Electrical Load} 
We consider the non-dispatchable electrical loads in the microgrids.
The demand of the electrical loads in microgrid $i$ during time slot $t$ is denoted as $l_i(t)$.
\subsection{Battery Charging and Discharging}
During a time slot, a microgrid may charge or discharge the battery for energy storage or utilization.
The changes of the battery level of the batteries in a microgrid due to the charging and discharging operations can be modeled as
\begin{equation}
b_i(t) = b_i(t-1) + c_i(t) \beta^c_i - \frac{d_i(t)}{\beta_i^d}, 0 < \beta_i^c, \beta_i^d < 1,
\end{equation}
where $\beta_i^c$ and $\beta_i^d$ are the charging and discharging efficiency, 
$c_i(t)$ is the amount of energy charged to the batteries in microgrid $i$ during time slot $t$,
and $d_i(t)$ is the amount of energy discharged from the batteries in microgrid $i$ during time slot $t$.
We assume that either charging or discharging can be performed to the batteries of a microgrid during a time slot.

\subsection{Energy Trading among Microgrids}
At the beginning of a time slot, each microgrid determines whether it should participate in the energy trading of the next time slot on the hour-ahead market to sell or to buy electricity.
Each microgrid determines the trading price and the trading quantity that will be submitted to the bidding system.
The final market clearing price and trading quantity will be determined based on the double auction mechanism.

\begin{figure}
\begin{center}
\epsfig{file=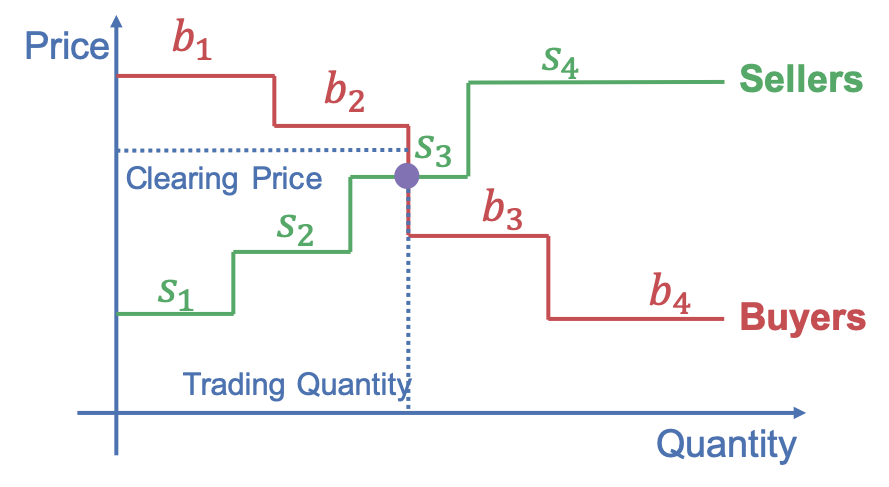, width=0.95\columnwidth}
\end{center}
\caption{The double auction mechanism. The buyers ($b_i$) will be sorted in decreasing order and the sellers ($s_i$) will be sorted in increasing order. The intersection point determines the clearing price and trading amount.} 
\label{fig:double-auction}
\end{figure}

\paragraph{Double Auction Mechanism}
We design a double auction mechanism suitable for multiagent reinforcement learning-based trading according to the method proposed in \cite{wang2014game}.
As illustrated in Fig. \ref{fig:double-auction},
the buyers' bid price at time slot $t$ will be sorted in decreasing order,  
denoted as $\Lambda_1(t) > \Lambda_2(t) > \Lambda_3(t) > ...$.
The sellers' ask prices at time slot $t$ will be sorted in increasing order, 
denoted as $\Gamma_1(t) < \Gamma_2(t) < \Gamma_3(t) < ...$.
The aggregate supply curve and demand curve will intersect at a point
where satisfies the following inequality, 
\begin{equation}
\Lambda_{k}(t) > \Gamma_{l}(t) > \Lambda_{k+1}(t),
\end{equation}
The market clearing price can be any value within $(\Gamma_l(t), \Lambda_k(t))$.
In this work, we adopt $p^*(t) = (\Lambda_k(t) + \Gamma_l(t))/2$ as the market clearing price at time slot $t$.
If the total demands of the $k$ buyers and the total supply of the $l$ sellers are mismatching,
the buyers with a higher bid price or the sellers with a lower ask price will be satisfied first.
If the market needs to guarantee the truthfulness of double auction,
it can be modified by excluding seller $k$ and buyer $l$ \cite{wang2014game}.

\paragraph{Monetary Cost for Buying Energy on the Hour-ahead Market}
If a microgrid successfully bid a certain quantity of energy on the hour-ahead market, it needs to make the payment to the market.
The monetary cost is determined by the market clearing price and trading quantity, which are determined by double auction.
Suppose microgrid $i$ bids successfully, the monetary cost for microgrid $i$ during time slot $t$ is
\begin{equation}
\rho_i(t) = p^*(t) \chi_i(t), 
\end{equation}
where $\chi_i(t)$ is the bid quantity of microgrid $i$ from the hour-ahead market at time slot $t$ and $p^*(t)$ is the clearing price.

\paragraph{Revenue for Selling Energy on the Hour-ahead Market}
If a microgrid successfully sell energy on the hour-ahead market, it can get a revenue, 
which is determined by the market clearing price and the trading quantity by double auction.
Note that the final energy supply from a microgrid to the energy delivery system may be less than the pre-determined trading quantity due to device outrage, bad weather condition or yield prediction error.
Thus, the revenue will be calculated based on the real energy supply from a microgrid. 
The revenue for microgrid $i$ on the hour-ahead market during time slot $t$ is
\begin{equation}
\sigma_i(t) = p^*(t) v_i(t),
\end{equation}
where $v_i(t)$ is the supplied energy from microgrid $i$ during time slot $t$.
Note that $v_i(t)$ should be no more than the trading quantity determined by double auction for microgrid $i$.

\paragraph{Penalty for Unfulfilling the Contract}
After the trading is completed, each microgrid needs to fulfill the contract. 
For the sellers, they must supply the gird with the amount of energy that is equal to the pre-determined trading quantity by double auction.
If a seller's energy supply to the grid is less than the pre-determined trading amount for any reasons,
the seller will be charged some penalty fees for violating the contract.
The penalty for microgrid $i$ is calculated as
\begin{equation}
q_i(t) = f(u_i(t) - v_{i}(t)), \ u_i(t) \ge v_{i}(t),
\end{equation}
where $q_i(t)$ is the penalty charged to microgrid $i$ at the end of time slot $t$, 
$u_i(t)$ is the pre-determined selling quantity of microgrid $i$ by double auction at the beginning of time slot $t$,
$v_{i}(t)$ is the real energy supply of microgrid $i$ to the grid during time slot $t$, and $f$ is the weight of penalty.

In this paper, we assume $f$ is the gap between the wholesale market price and the market clearing price at time slot $t$.
The practical meaning is that some buyers will utilize $u_i(t)$ units of energy during time slot $t$ at the market clearing price $p^*(t)$ by trading;
if microgrid $i$ only supply $v_{i}(t)$ units, the balance will be made up by the wholesale market,
and microgrid $i$ will be penalized based on the difference between the market clearing price and the wholesale market price at time slot $t$.

\paragraph{Monetary Cost for Buying Energy from the Wholesale Market}

If a microgrid consumes more energy from the grid than its buy quantity from the hour-ahead bidding market,
the exceeding part of the energy consumption will be charged based on the price of the wholesale market.
The overall supply of energy to a microgrid and its overall consumption of energy should be balanced during a time slot.
Based on this balance, we can calculate the consumed energy from the wholesale energy market by microgrid $i$  during time slot $t$ as
\begin{equation}
\varpi_i(t) = c_i(t) + v_i(t) + l_i(t) - g_i(t) - d_i(t) - \chi_i(t).
\end{equation}

Note that only one type of operation can be conducted for a microgrid during a time slot for charging and discharging, selling and buying, at least one of the two variables should be zero. 
The monetary cost for the consumed energy from the wholesale market by microgrid $i$ during time slot $t$ is
\begin{equation}
\xi_i(t) = p_w \varpi_i(t),
\end{equation}
where $p_w$ is the electricity price of the wholesale market.

\subsection{Problem Formulation}
We formulate the energy trading and scheduling among microgrids as a Markov Game with continuous action spaces.
Each microgrid is an agent, which aim is to learn the optimal policy to maximize its overall rewards.

\paragraph{State}
The state of each microgrid for each time slot consists of two parts, namely, the local state and the public state.
The local state includes the current battery level, 
the historical energy generation power and the historical electrical load of a microgrid.
The public state includes the price of the wholesale market and the historical market clearing price in the hour-ahead market.
Because the energy generation power and the electrical load for time slot $t$ are unknown, we adopt the information of last time slot $t-1$ to represent the states.
We denote the state of microgrid $i$ at time slot $t$ as
\begin{equation}
s_i(t) = (b_i(t), g_i(t-1), l_i(t-1), p_w, p^*(t-1)),
\end{equation}
where $b_i(t)$ is the battery level at the beginning of time slot $t$, $g_i(t-1)$ is the renewable energy generation power during time slot $t-1$,
$l_i(t-1)$ is the load during time slot $t-1$,
$p_w$ is the wholesale market price, and $p^*(t-1)$ is the market clearing price of the hour-ahead market at time slot $t-1$.

\paragraph{Action}
The control action for each microgrid during each time slot $t$ includes the selling/buying price, the selling/buying quantity, and the charging/discharging quantity.
We denote the control action for microgrid $i$ at time slot $t$ as
\begin{equation}
a_i(t) = (\Lambda_i(t)|\Gamma_i(t), \phi_i(t)|\lambda_i(t), c_i(t)|d_i(t)),
\end{equation}
where operator $|$ represents choosing one action out of the two,
$\Lambda_i(t)/\Gamma_i(t)$ are the submitted price of microgrid $i$ for selling/buying energy at time slot $t$ on the hour-ahead market,
$\phi_i(t)/\lambda_i(t)$ are the submitted selling/buying quantity of microgrid $i$ at time slot $t$,
$c_i(t)/d_i(t)$ are the charging/discharging quantity of microgrid $i$ during time slot $t$. 
Note that a microgrid can be either a seller or a buyer during a time slot,
and it can either charge battery or discharge battery.
In the implementation, we can use the different ranges of an output of a neural network to represent different operations.
For instance, the positive numbers of an output represent charging and the negative numbers represent discharging.
Similarly, the positive numbers of an output represent selling quantity and the negative numbers represent buying quantity.

\paragraph{Reward}
The reward for a microgrid during a time slot is the summation of the revenue and cost, which consist of four parts: 
1) the revenue for selling energy on the hour-ahead market $\sigma_i(t)$,
2) the cost for buying energy from the wholesale market $\xi_i(t)$,
3) the cost for buying energy from the hour-ahead market $\rho_i(t)$,
4) the penalty for energy under-supply to the grid $q_i(t)$.
We define the reward function of microgrid $i$ for time slot $t$ as follow,
\begin{equation} \label{eqn:reward_function}
r_i(t) = \sigma_i(t) - \xi_i(t) - \rho_i(t) - q_i(t).
\end{equation}
The reward reflects the quality of the control policy and each microgrid aims to maximize its reward.

\paragraph{Optimization Objective}
The problem of energy trading and scheduling  among microgrids can be formulated as a Markov Game \cite{littman1994markov} with $N$ agents and continuous action spaces.
We denote a Markov Game as a tuple, $<\mathcal{S}, N, \mathcal{A}_i, \mathcal{T}, \mathcal{R}_i>$,
where $\mathcal{S}$ is the set of states describing the environments of all agents,
$N$ is the number of agents,
$\mathcal{A}_i$ is the set of actions of agent $i$,
$\mathcal{T}$ is the state transition function of the environment,
$\mathcal{R}_i$ is the set of rewards of agent $i$. 
At each time slot $t$, the agents observe their local states $s_1(t), s_2(t),..., s_N(t)$ and take control actions
$a_1(t), a_2(t), ..., a_N(t)$ according their control policies $\pi_1, \pi_2, ..., \pi_N$.
At the end of time slot $t$, the agents will receive their rewards $r_1(t), r_2(t), ..., r_N(t)$.
The reward for an agent is determined by the state of the whole environment and the actions of all agents.
The states of the agents will evolve to new states $s_1(t+1), s_2(t+1), ..., s_N(t+1)$ according to the state transition probability function.
Our aim is to derive the optimal control policy for each agent to maximize its overall discounted future rewards,
which can be presented as
\begin{equation}
\pi^{*}_i = \argmax_{\pi} \mathop{\mathbb{E_\pi}} \sum_{k=0}^{\infty} \gamma^{k}_i r_i(t+k),
\end{equation}
where $\pi^*_i$ is the optimal control policy for microgrid $i$,
$\gamma_i$ is the discount factor for the reward of agent $i$,
and $r_i(t+k)$ is the reward for agent $i$ at time slot $t+k$. 
The state transition probability function $\mathcal{T}$ is hard to obtain for a complex environment, 
we will introduce the multiagent deep reinforcement learning approach to learn the optimal policy for each agent by directly interacting with the environment.

\section{Learning Algorithm for Distributed Energy Trading and Scheduling} \label{sec:algorihtm}
In this section, we introduce the algorithm for learning the optimal control policy for each microgrid.

\subsection{Choice of Algorithm}
We first briefly introduce why we choose MADDPG for learning the optimal control policy for distributed energy trading and scheduling among microgrids.

\paragraph{Mixed Cooperative-Competitive}
Each autonomous microgrid can be seen an agent, which makes the decision for maximizing its rewards. 
The relationships among the interconnected microgrids are mixed cooperative-competitive.
One microgrid may compete with other microgrids for energy selling on the hour-ahead market to maximize revenue;
On the other hand, it could also buy energy from other microgrids at a lower price compared with the wholesale market to reduce cost.
MADDPG can learn the optimal policy for each agent in a mixed cooperative-competitive environment.

\paragraph{Decentralized Control}
MADDPG adopts the framework of centralized training and decentralized execution.
During the training, some extra information of other agents (e.g., action, rewards, and training episodes) will be used for training the policy of an agent for learning collaboratively.
However, the private information of other microgrids will not be used during execution,
and only the local information of a microgrid and the public information are used for local decision making.
Thus, MADDPG is suitable for decentralized control to maintain the autonomy of each microgrid.

\paragraph{Continuous Control}
Most control actions in a microgrid are continuous, e.g., charging/discharging quantity, trading price and trading quantity, etc. 
MADDPG is a multiagent extension of Deep Deterministic Policy Gradient (DDPG), which is a deep reinforcement learning algorithm for solving continuous control problem.
Therefore, MADDPG is naturally suitable for addressing the continuous control problem.

\subsection{Design of Actor and Critic Networks} 
\begin{figure}
\begin{center}
\epsfig{file=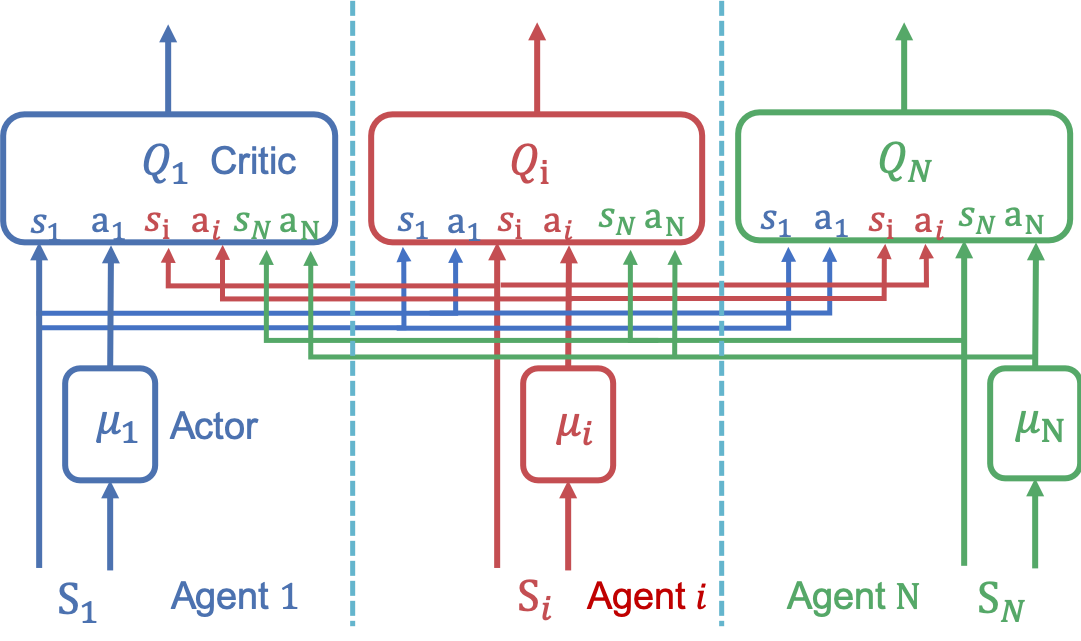, width=0.9\columnwidth}
\end{center}
\caption{The network architecture of MADDPG. The actor of an agent inputs the local state of a microgrid and outputs control action. The critic of an agent inputs the states and actions of all agents and outputs an evaluation.} 
\label{fig:maddpg-network}
\end{figure}
We illustrate the network architecture of MADDPG in Fig. \ref{fig:maddpg-network}.
Each agent consists of an actor and a critic.
The actor maps the state to the control action.
The critic evaluates the advantage of the actions under the given states compared to the average actions.
We denote the actor network of agent $i$ as $\mu_i(a_i|s_i)$, where $s_i$ is the state of microgrid $i$ and $a_i$ is the corresponding control action for microgrid $i$.
The actor network of an agent inputs the local state and outputs a control action.
We denote the critic network of agent $i$ as $Q_i(s_1,a_1,...,s_i,a_i,...,s_N,a_N)$.  
The critic network inputs the states $(s_1, s_2, ..., s_N)$ and the actions $(a_1, a_2, ..., a_N)$ of all agents and outputs a real-valued evaluation.

Each agent has an individual critic network, 
which allows the agents to have different reward structures,
and a microgrid can learn its optimal policy based on its objectives in a cooperative, competitive, or mixed environment.
We only use the critic networks for training the actor network of each agent.
After training, the critic networks are no longer required, and we only use the actor network of an agent for making control action for the corresponding microgrid. 
Thus, it will not relay on the states and actions of other microgrids.

\begin{figure}
\begin{center}
\epsfig{file=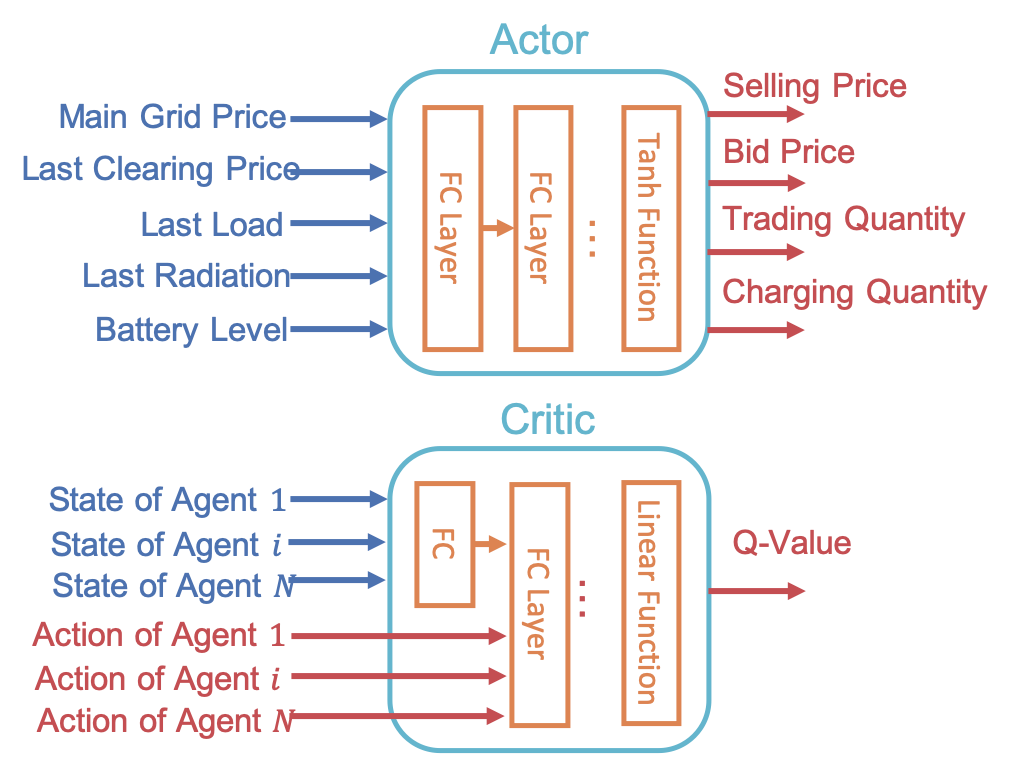, width=0.85\columnwidth}
\end{center}
\caption{The implementation of the actor network and the critic network of each agent. If output value of the trading quantity is larger than zero, the agent will be a buyer. If less than zero, the agent will be a seller.} 
\label{fig:maddpg-implementation}
\end{figure}

We illustrate the implementation of the actor network and the critic network in Fig. \ref{fig:maddpg-implementation}.
The outputs of an actor network include the selling price, the bid price, the trading quantity, and the charging quantity.
Because an agent can only be either a seller or a buyer within one time slot, we use the value of the output of trading quantity to determine its role.
Specifically, if the output value of trading quantity is larger than zero, the agent will be a buyer, and the bid price and the trading quantity will be submitted to the hour-ahead market for biding.
If the output value of the trading quantity is less than zero, the agent will be a seller, and the selling price and trading quantity will be submitted for bidding on the hour-ahead market.

In the critic network of an agent, the states of all agents will be input into a Fully-Connected (FC) layer.
The output vector of the FC layer will be concentrated with the control actions of all actors, and then input the second FC layer.
Note that the main grid price and the last clearing price are public information, which are same for each agent. Therefore, they will be input into the critic network of an agent only once to avoid duplication and reduce the input dimension of the critic networks.
The activation function of the actor network is Tanh function, and the range of each output is from $-1$ to $1$,
and we will convert them to the actual range of each control action.
The activation function of the critic network is a linear function and the output is a real value. 

\subsection{Learning Algorithm}
We now describe how to train the actor network and the critic network of each agent.
The learning process can be conducted in a simulation environment.
The details of the training algorithm are illustrated in Algorithm \ref{algo:training-algo}.

At the beginning of each time slot $t$, each agent $i$ first observes the state $s_i$ of microgrid $i$. 
State $s_i$ will be input into the actor network $\mu_i$ of agent $i$ and output a control action.
For the exploration of the state space, we will add an exploration noise to the control action, and the final control action is  
\begin{equation} \label{eqn:control-action}
a_i = \mu_{i}(s_i) + \mathcal{N},
\end{equation}
where $\mathcal{N}$ is the exploration noise.
We adopt an Ornstein-Uhlenbeck process \cite{uhlenbeck1930theory} to generate the exploration noise.

After obtaining the control action for each agent $i$, the control actions will be applied in the corresponding microgrids. 
The simulation environment will simulate the energy scheduling process and the energy trading process according to the specified control action for each microgrid. 
At the end of a time slot, 
each agent $i$ can calculate its reward $r_i$ during the time slot by Eq. \eqref{eqn:reward_function} and observe new state $s'_i$.
The state, action, reward, and new state information of each agent during a time slot will be stored as a transition in the replay buffer as $(\bold{s},\bold{a},\bold{r},\bold{s'})$,
where $\bold{s}=(s_1,...,s_N)$, $\bold{a}=(a_1,...,a_N)$, $\bold{r}=(r_1,...,r_N)$, and $\bold{s'}=(s'_1,...,s'_N)$.

At the end of each time slot, the actor network and the critic network of each agent will be trained with the transitions sampled from the replay buffer.
To stabilize training, a copy of the actor networks and the critic networks will be created as target networks for slowly tracking the learned networks.
We denote the target actor network and the target critic network of agent $i$ as $\mu'_i$ and $Q'_i$, respectively. 
To train the actor network and the critic network of an agent, we randomly select $S$ transitions from the replay buffer. 
For transition $j$, we denote it as $(\bold{s}^j, \bold{a}^j, \bold{r}^j, \bold{s'}^{j})$.
The target Q-value of this transition for agent $i$ will be calculated as follow,
\begin{equation}\label{eqn:target-value}
y^j_i = r^j_i + \gamma Q'_i(\bold{s}'^j, a'_1,...,a'_{N})|_{a'_k=\mu'_k(s'_k), k=1,...,N,}
\end{equation}
where $\gamma$ is the discounting factor and $s'_k$ is the $k$-th item in $\bold{s'}^{j}$ (i.e., the next state of agent $k$).
The critic network of agent $i$ will be updated by minimizing the following loss,
\begin{equation} \label{eqn:loss-function}
L(\theta_i) = \frac{1}{S} \sum_{j=1}^S(y^j_i-Q_i(\bold{s}^j,\bold{a}^j))^2,
\end{equation}
where $\theta_i$ denotes the network parameters of agent $i$.
The actor network of agent $i$ will be updated using the policy gradient (see details in \cite{lowe2017multi}).
%
%
%
The target network parameters of each agent $i$ will be updated using the following equation.
\begin{equation} \label{eqn:update-target-network}
\theta'_i  \gets \tau \theta_i + (1-\tau)\theta'_i,
\end{equation}
where $\theta'_i$ represents the network parameters of the target network of agent $i$ and $\tau$ is the learning rate.

We denote the actor network of agent $i$ after training is $\mu_i^*$.
The control action $a_i$ for microgrid $i$ can be directly obtained by observing the state $s_i$ of microgrid $i$ and inputing $s_i$ into the actor network, mathematically denoted as $a_i = \mu_i^*(s_i)$. 
Thus, the computational complexity for obtaining a control action is only determined by the complexity of the actor network, which include the number of hidden layers and the number of neurons in each hidden layer.
In this work, the actor network of an agent consists of two hidden layers. 
We suppose that each hidden layer of the actor has $n$ neurons, the complexity for making a control action for an agent is $O(n^2)$.

\begin{algorithm}
\renewcommand{\algorithmicrequire}{\textbf{Input:}}
\renewcommand\algorithmicensure {\textbf{Output:} }
\caption{Training Algorithm for MADDPG} \label{algo:training-algo}
\begin{algorithmic}[1]
%
\FOR{episode = 1, 2, ..., M}
\STATE{Observe the microgrids' states, $\bold{s}=(s_1,...,s_N)$}
\FOR{step t = 1, 2, ..., T}
\STATE{Make control action for each agent $i$} by Eq. \eqref{eqn:control-action}.
\STATE{Apply $\bold{a}=(a_1,...,a_N)$ in the corresponding microgrids for energy trading and scheduling}
\STATE{Calculate reward $\bold{r}=(r_1,...,r_N)$ by Eq. \eqref{eqn:reward_function}}
\STATE{Observe the microgrids' new state $\bold{s'}=(s'_1,...,s'_N)$}
\STATE{Store transition $(\bold{s},\bold{a},\bold{r},\bold{s'})$ into the replay buffer}
\STATE{Set $\bold{s} \leftarrow \bold{s'}$}
\FOR{agent $i = 1,2,...,N$}
\STATE{Sample $S$ transitions from the replay buffer}
\STATE{Calculate $y^j_i$ for each transition by Eq. \eqref{eqn:target-value}}
\STATE{Update critic of agent $i$ by minimizing Eq. \eqref{eqn:loss-function}}
\STATE{Update actor of agent $i$ using the policy gradient.} 
\STATE{Update target network of agent $i$ by Eq. \eqref{eqn:update-target-network}}
\ENDFOR
\ENDFOR
\ENDFOR
\end{algorithmic}
\end{algorithm}

\begin{figure*}
\centering
\subfigure[Microgrid 1]{\label{fig:a}\includegraphics[width=0.5\columnwidth]{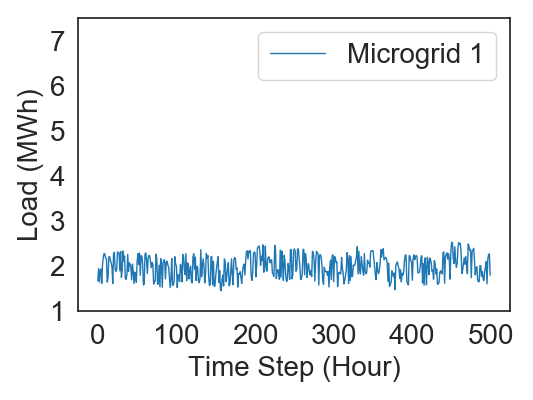}}
\subfigure[Microgrid 2]{\label{fig:b}\includegraphics[width=0.5\columnwidth]{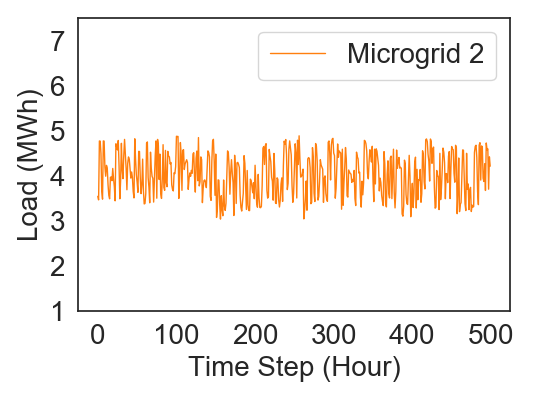}}
\subfigure[Microgrid 3]{\label{fig:a}\includegraphics[width=0.5\columnwidth]{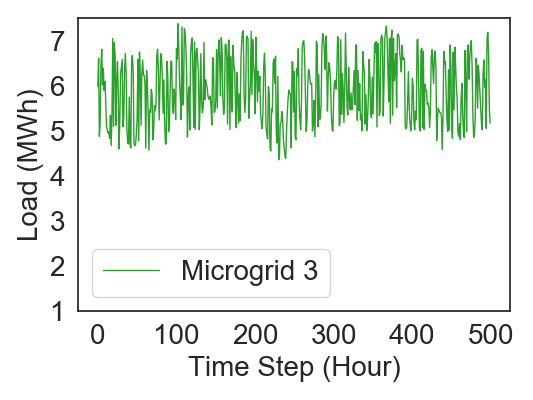}}
\subfigure[Microgrid 4]{\label{fig:a}\includegraphics[width=0.5\columnwidth]{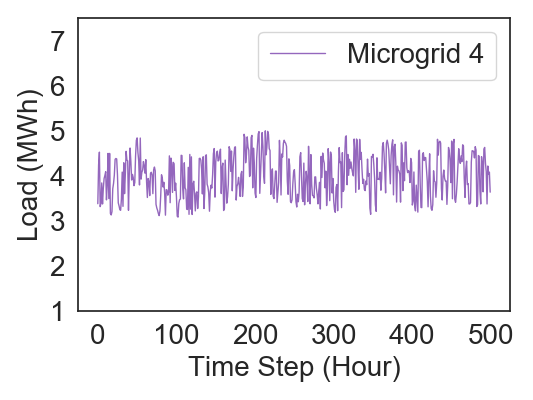}}
\caption{The load profile of the four microgrids. The microgrids have different electrical load characteristics, and the electrical loads are time-varying.}
\label{fig:load_profile}
\end{figure*}

\begin{figure*}
\centering
\subfigure[Microgrid 1]{\label{fig:a}\includegraphics[width=0.5\columnwidth]{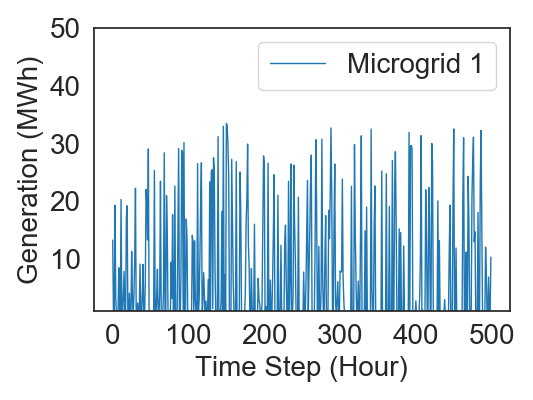}}
\subfigure[Microgrid 2]{\label{fig:b}\includegraphics[width=0.5\columnwidth]{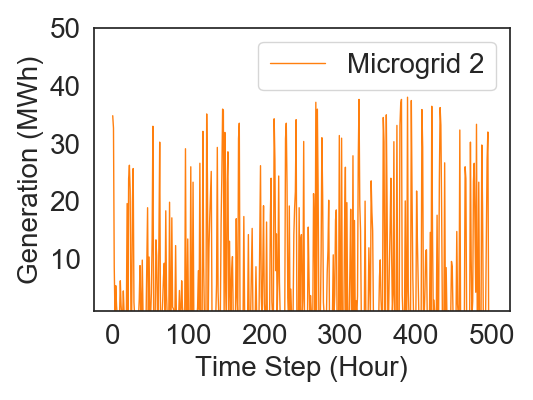}}
\subfigure[Microgrid 3]{\label{fig:a}\includegraphics[width=0.5\columnwidth]{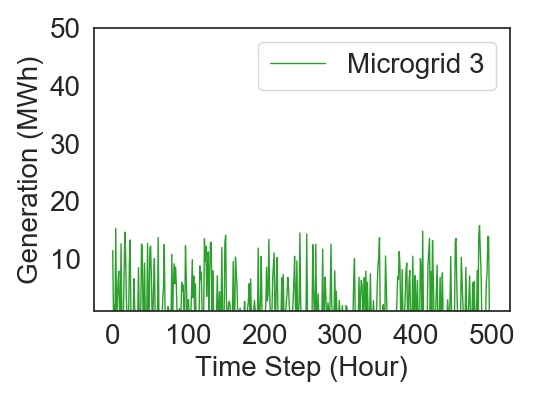}}
\subfigure[Microgrid 4]{\label{fig:a}\includegraphics[width=0.5\columnwidth]{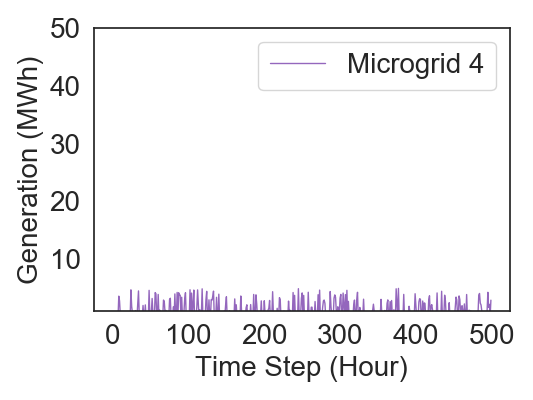}}
\caption{The renewable energy generation of each microgrid. Microgrid 1 and 2 have more renewable energy generations and Microgrid 3 and 4 have less.}
\label{fig:solar_profile}
\end{figure*}

\begin{figure*}
\centering
\subfigure[Microgrid 1]{\label{fig:a}\includegraphics[width=0.5\columnwidth]{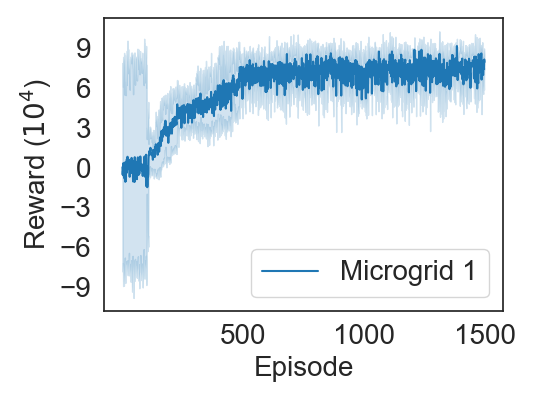}}
\subfigure[Microgrid 2]{\label{fig:b}\includegraphics[width=0.5\columnwidth]{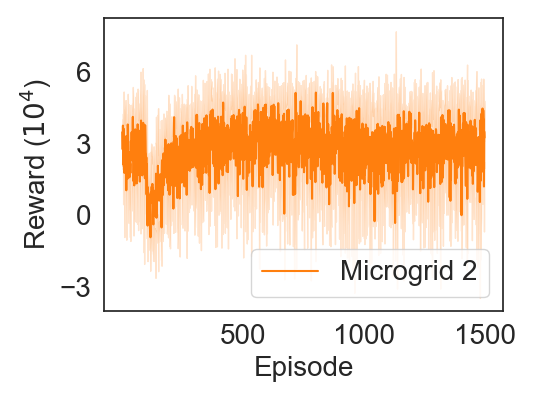}}
\subfigure[Microgrid 3]{\label{fig:a}\includegraphics[width=0.5\columnwidth]{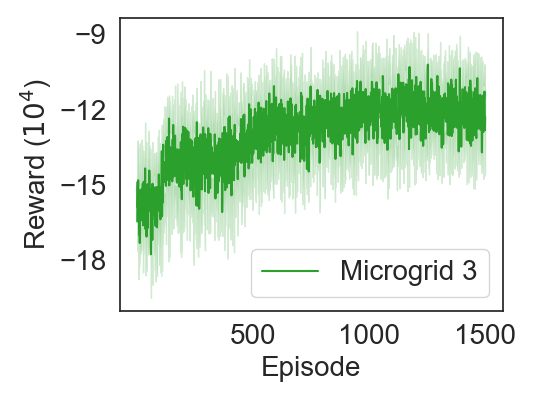}}
\subfigure[Microgrid 4]{\label{fig:a}\includegraphics[width=0.5\columnwidth]{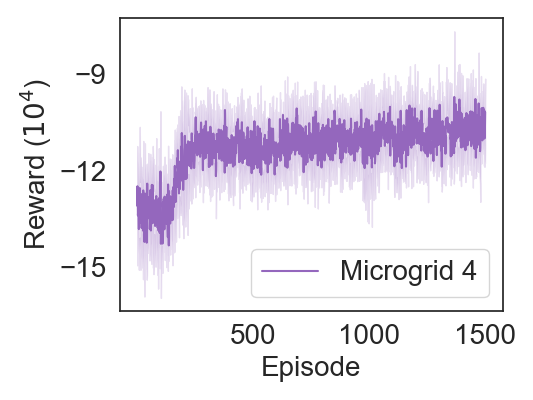}}
\caption{The convergences of the rewards of different agents. The rewards of the agents will converge after training.}
\label{fig:reward_convergence}
\end{figure*}

\section{Experiment} \label{sec:performance}
In this section, we introduce the experiment settings and evaluate the performances of our proposed method.

\subsection{Experiment Setting and Datasets}
We simulate the energy trading and scheduling among four microgrids with different load characteristics and renewable energy generation capacities. 
The electrical load and the renewable energy generation of each microgrid during each time slot are scaled from real-world datasets.
We illustrate parts (500 hours) of the load profiles and the renewable energy generation profiles of each microgrid in Fig. \ref{fig:load_profile} and Fig. \ref{fig:solar_profile}, respectively.
Microgrid 1 and 2 have more renewable energy generations and surplus energy to sell.
Microgrid 3 and 4 have heavier loads and less renewable energy generations.

The electricity price on the wholesale market is 22.79 cents/kWh in Singapore. 
The minimum bidding price on the hour-ahead market is 15.0 cents/kWh, and the maximum bidding price is 22.79 cents/kWh, which is equal to the wholesale market.
We use the solar radiation data of one year at four locations of Singapore, which are obtained from Solcast \cite{solcast}, to simulate the solar radiation in each microgrid.
We use the scaled historical electricity system demand data for every half-hour period in Singapore to simulate the electrical loads of the microgrids \cite{load_sg}. 
The maximum battery capacities of Microgrid 1-4 are 100MWh, 100MWh, 20MWh, 10MWh, respectively.
The maximum bid quantity for each microgrid on the hour-ahead market is 7.5MWh.

The initial exploration noise scale is 1.0, and the exploration noise linearly decreases over 1000 training episodes till zero.
The models are trained over 1500 episodes, and the performances are evaluated over 1000 episodes.
Each episode consists of $7*24$ time steps, and each time step in the simulation is one hour.
The batch size is 1024. The learning rate is 0.01.
The discount factor of the reward is 0.8.
Each critic network has three hidden layers, the first layer has 1024 neurons, the second layer has 512 neurons, and the third layer has 256 neurons.
Each actor network has two hidden layers, the first layer has 512 neurons, and the second layer has 128 neurons.
The activation function is ReLU function.

\subsection{Convergence Analysis}
To evaluate the convergences of different agents in MADDPG, we train the models five times under the same settings with different random seeds.
We illustrate the rewards of different agents at different training episodes in Fig. \ref{fig:reward_convergence}, respectively. 
The shadow areas in the figures represent the standard deviations of the rewards of the five training times at the different episodes.
We can observe from these figures that the rewards of different agents all converge.

The converged rewards of Microgrid 1 and 2 are larger than zero.
This is because Microgrid 1 and 2 have larger capacities for renewable energy generation than their own consumptions.
Therefore, they can sell their surplus energy on the hour-ahead market and gain revenues.
Their revenues are larger than their cost, therefore, their converged rewards are larger than zero.
The rewards of Microgrid 3 and 4 also increase during the training,
because they can save more cost by improving the trading policy on the hour-ahead market.
By trading on the hour-ahead market, they can save more cost compared with buying energy from the wholesale market.

The results illustrated in Fig. \ref{fig:reward_convergence} verify the feasibility of adopting the multiagent approach for learning the optimal control policy for each microgrid.
The four microgrids in the studied case are mixed cooperative-competitive.
Microgrid 1 and 2 are competitors for energy selling, and Microgrid 3 and 4 are competitors for energy buying.
Microgrid 1 and 2 also collaborate with Microgrid 3 and 4 for energy trading for mutual benefit.
The convergences of the agents verify that the agents can find the equilibrium in the mixed cooperative-competitive environment for maximizing their rewards.

\begin{figure*}
\centering
\subfigure[Microgrid 1]{\label{fig:a}\includegraphics[width=0.5\columnwidth]{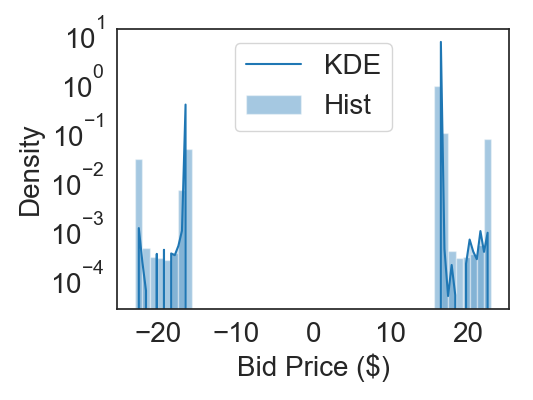}}
\subfigure[Microgrid 2]{\label{fig:b}\includegraphics[width=0.5\columnwidth]{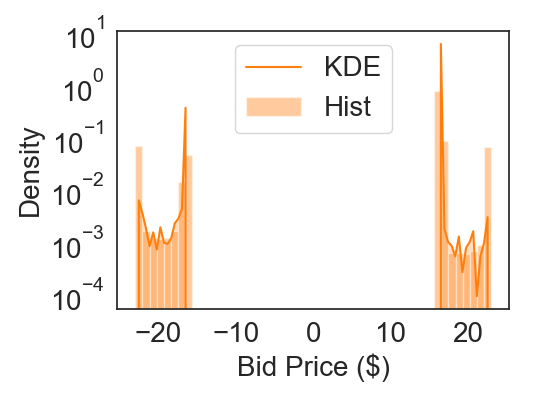}}
\subfigure[Microgrid 3]{\label{fig:a}\includegraphics[width=0.5\columnwidth]{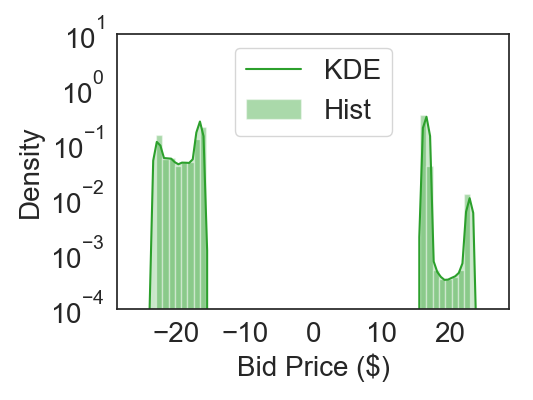}}
\subfigure[Microgrid 4]{\label{fig:a}\includegraphics[width=0.5\columnwidth]{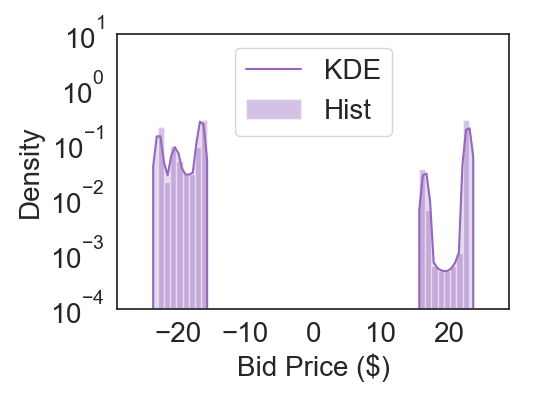}}
\caption{The distribution of the bid price of different microgrids on the hour-ahead market. }
\label{fig:bid_price}
\end{figure*}

\begin{figure*}
\centering
\subfigure[Microgrid 1]{\label{fig:a}\includegraphics[width=0.5\columnwidth]{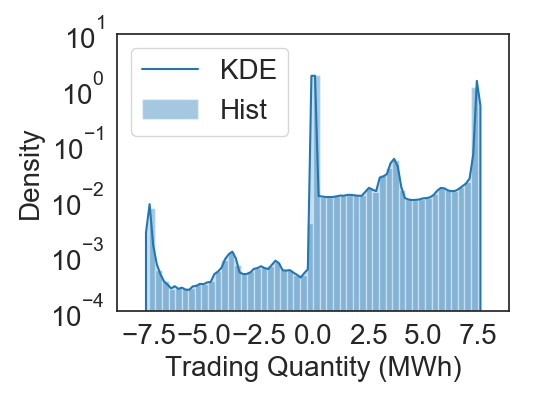}}
\subfigure[Microgrid 2]{\label{fig:b}\includegraphics[width=0.5\columnwidth]{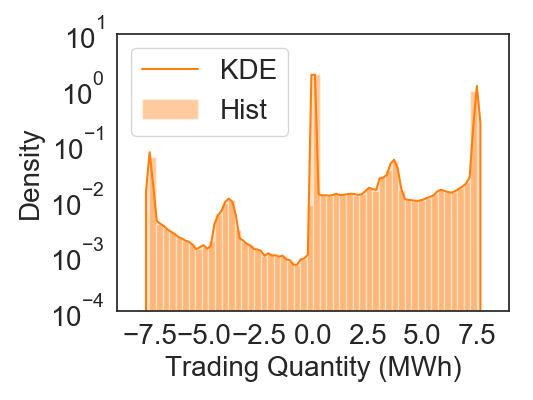}}
\subfigure[Microgrid 3]{\label{fig:a}\includegraphics[width=0.5\columnwidth]{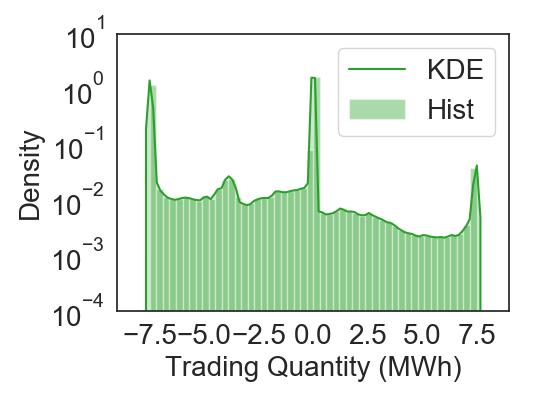}}
\subfigure[Microgrid 4]{\label{fig:a}\includegraphics[width=0.5\columnwidth]{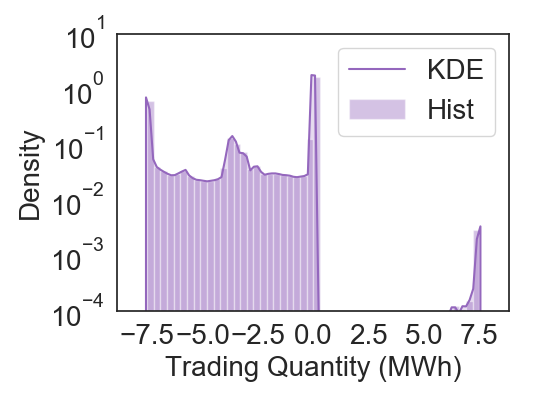}}
\caption{The distribution of the trading quantity of different microgrids on the hour-ahead market.}
\label{fig:bid_quantity}
\end{figure*}

\begin{figure*}
\centering
\subfigure[Microgrid 1]{\label{fig:a}\includegraphics[width=0.5\columnwidth]{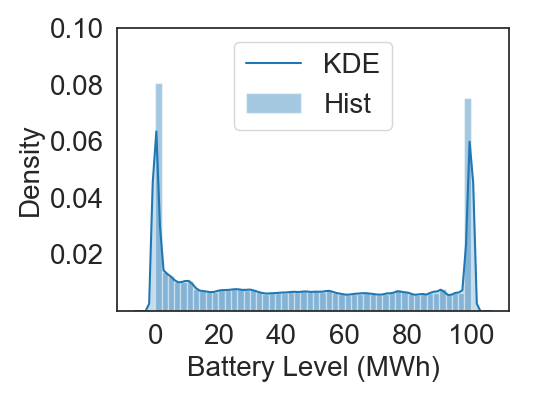}}
\subfigure[Microgrid 2]{\label{fig:b}\includegraphics[width=0.5\columnwidth]{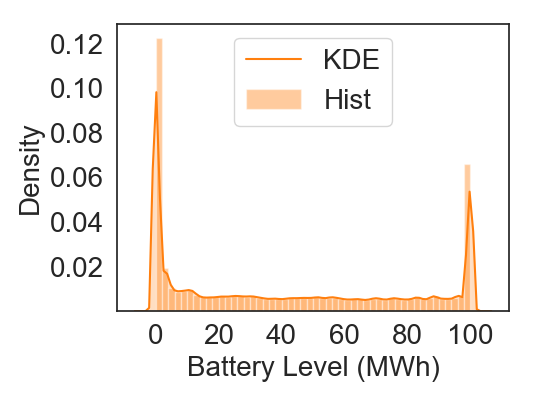}}
\subfigure[Microgrid 3]{\label{fig:a}\includegraphics[width=0.5\columnwidth]{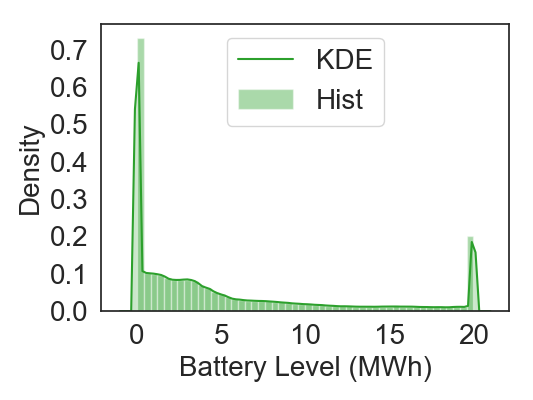}}
\subfigure[Microgrid 4]{\label{fig:a}\includegraphics[width=0.5\columnwidth]{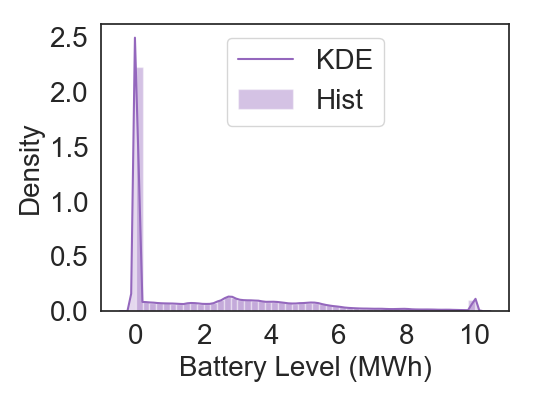}}
\caption{The distribution of the battery levels of different microgrids during different time slots.}
\label{fig:battery_state}
\end{figure*}

\begin{figure}
\begin{center}
\epsfig{file=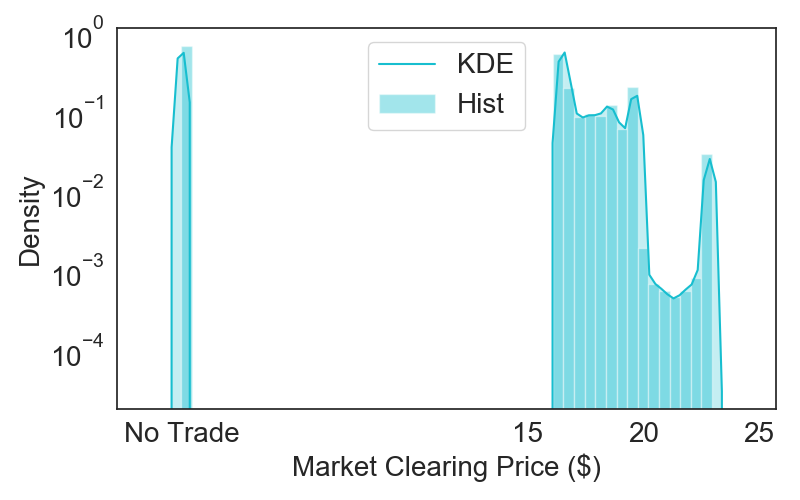, width=0.7\columnwidth}
\end{center}
\caption{The distribution of clearing price on the hour-ahead market.} 
\label{fig:clearing_price}
\end{figure}

\begin{figure}
\begin{center}
\epsfig{file=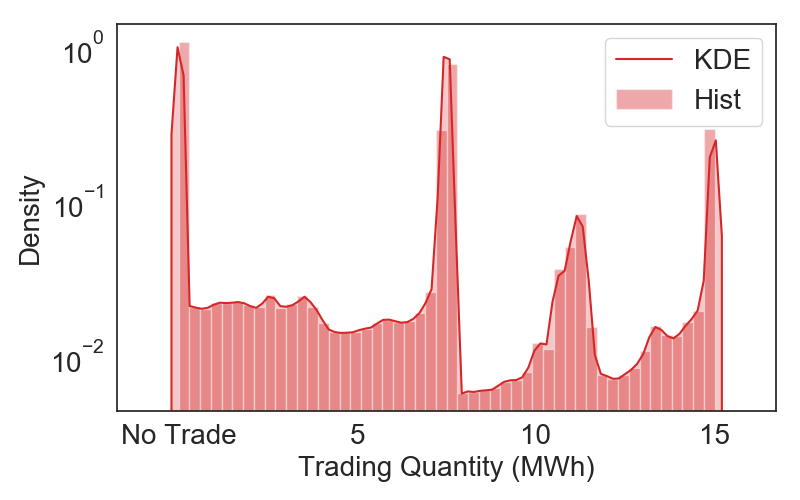, width=0.7\columnwidth}
\end{center}
\caption{The distribution of overall trading quantity on the hour-ahead market.} 
\label{fig:overall-trading-quantity}
\end{figure}

\subsection{Performance Analysis}
We analyze each agent's performances by illustrating the characteristics of the system states and the control actions of each microgrid.
We collect the control actions and the system states during each time slot over 1000 testing episodes.
The distributions of the system states and the control actions are estimated with Kernel Density Estimation (KDE) and Histogram (Hist) to analyze the characteristics of each microgrid.

We illustrate the distribution of the bid price and the trading quantity of different microgrids on the hour-ahead market in Fig. \ref{fig:bid_price} and \ref{fig:bid_quantity}, respectively.
In our definition of the control action of each agent,
if the price of an agent during one time step is larger than zero,
the agent is a seller during the time slot.
If the price is less than zero, the agent is a buyer during the time slot. 
In Fig. \ref{fig:bid_price} and \ref{fig:bid_quantity}, we can observe that Microgrid 1 and 2 are mainly as the sellers on the hour-ahead market,
because they have more surplus energy, which incentives them to sell to obtain revenues.
Microgrid 3 and 4 are mainly as buyers on the hour-ahead market due to lower capacities of energy generation.
The distributions of bid price and trading quantity are in line with the characteristics of each microgrid's energy consumption and generation capacity.%

The distributions of the selling price of Microgrid 1 and 2 are sharp, while the buying price of Microgrid 3 and 4 are flatter,
because Microgrid 1 and 2 are more competitive on energy selling,
and the renewable energy may be wasted if batteries have been fully charged and the generation exceeds the consumption.
On the contrary, Microgrid 3 and 4 may change buying price based on their varying demands.

The distribution of the market clearing price on the hour-ahead market is illustrated in Fig. \ref{fig:clearing_price}.
The market clearing price mostly lies within the range (15, 20), 
which is much lower than the wholesale market price.
Therefore, Microgrid 3 and 4 have the incentives of buying from the hour-ahead market to reduce cost.
In some cases, there is no trade made on the hour-ahead market because the selling price is too high or the buying price is too low.

We illustrate the distribution of the overall trading quantity during a time slot on the hour-ahead market in Fig. \ref{fig:overall-trading-quantity}.
The microgrids prefer to submit the trading quantity at the maximum allowed quantity,
because Microgrid 3 and 4 can always save cost by buying more from the hour-ahead market if they can consume.
On the contrary, Microgrid 1 and 2 can get more revenues by selling more, 
if the trading quantity does not exceed the surplus.
Therefore, the trading quantity of a microgrid has a large probability of being on the threshold of the maximum allowed trading quantity.

\begin{figure}
\begin{center}
\epsfig{file=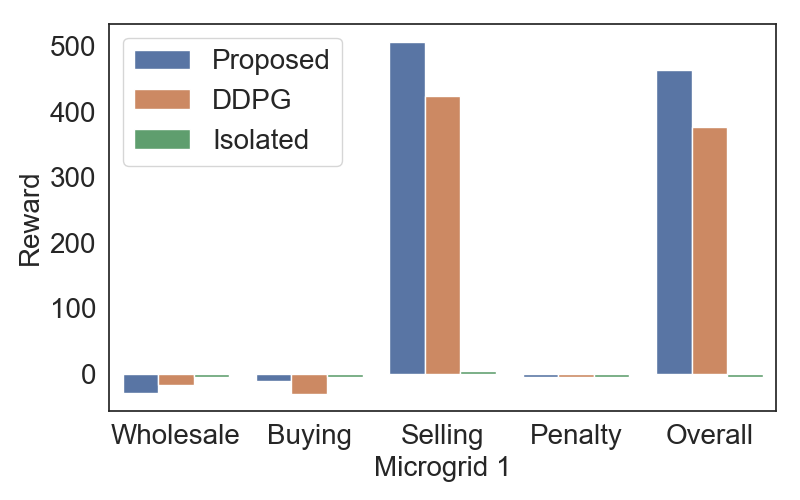, width=0.7\columnwidth}
\end{center}
\caption{The comparisons of the cost, revenue and the overall reward for Microgrid 1 under different methods.} 
\label{fig:cost_type_mg_1}
\end{figure}

\begin{figure}
\begin{center}
\epsfig{file=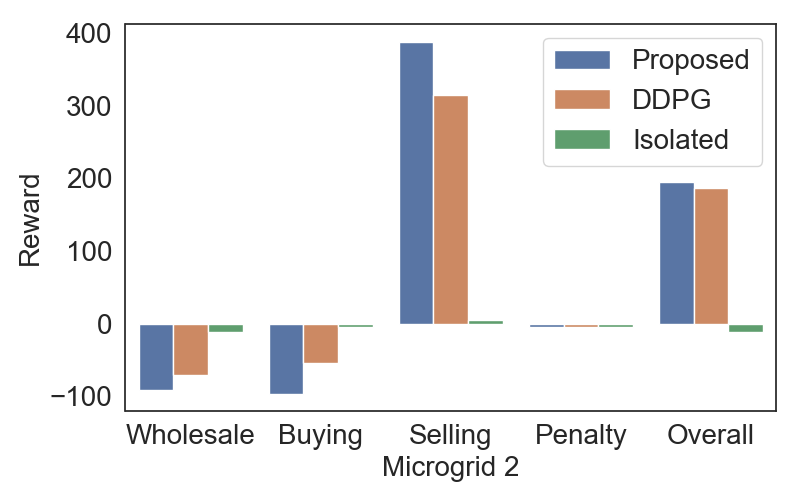, width=0.7\columnwidth}
\end{center}
\caption{The comparisons of the cost, revenue and the overall reward for Microgrid 2 under different methods.} 
\label{fig:cost_type_mg_2}
\end{figure}

\begin{figure}
\begin{center}
\epsfig{file=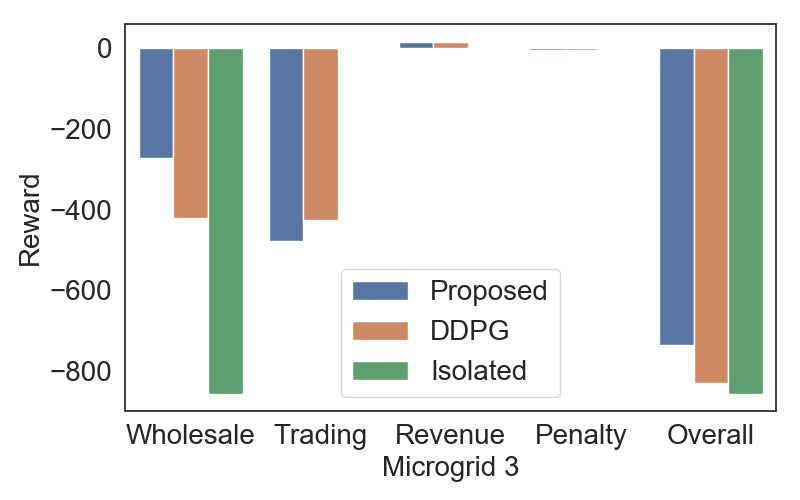, width=0.7\columnwidth}
\end{center}
\caption{The comparisons of the cost, revenue and the overall reward for Microgrid 3 under different methods.} 
\label{fig:cost_type_mg_3}
\end{figure}

\begin{figure}
\begin{center}
\epsfig{file=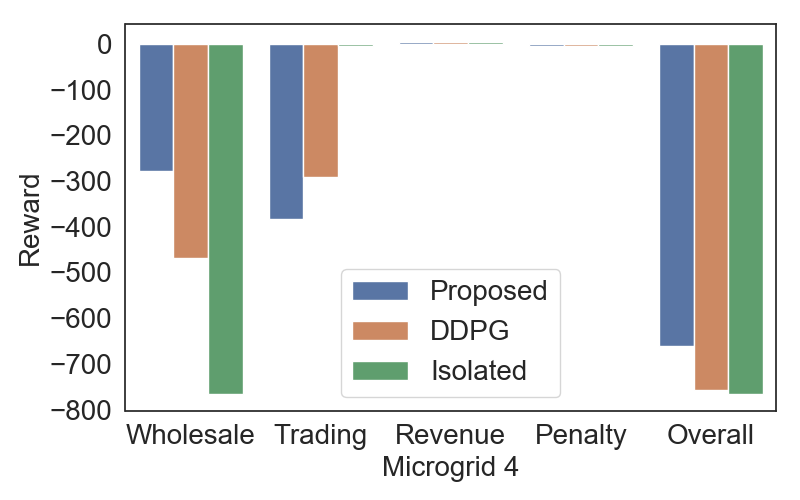, width=0.7\columnwidth}
\end{center}
\caption{The comparisons of the cost, revenue and the overall reward for Microgrid 4 under different methods.} 
\label{fig:cost_type_mg_4}
\end{figure}

We illustrate the distribution of the battery levels of different microgrids in Fig. \ref{fig:battery_state}.
The battery levels of Microgrid 3 and 4 are low during most of the time slots, because they have low capacity of energy generations.
For  Microgrid 1 and 2, their batteries may be fully charged, if they have too much surplus energy and cannot sell on the energy market,
which incentives them to sell on the hour-ahead market.
Therefore, the battery capacity is more important for the microgrids with larger capacity of renewable generation.

\subsection{Performance Comparison}
We compare the performances of our proposed method with the following baseline methods.

1) Isolated: the microgrids operate in an isolated mode, in which each microgrid can only supply the generated renewable energy to itself or charge the surplus energy to its battery.
A microgrid can only buy energy on the wholesale market to meet its demand when local generation is deficient.

2) DDPG: the microgrids can trade with each other, yet the agent of each microgrid is trained independently with DDPG, which is a single-agent reinforcement learning approach.
The agents will not share information with each other.

We illustrate the comparisons of the cost, revenue, and the overall reward per time slot for Microgrid 1-4 in Fig. \ref{fig:cost_type_mg_1}, \ref{fig:cost_type_mg_2}, \ref{fig:cost_type_mg_3}, \ref{fig:cost_type_mg_4}, respectively.
In the X-axises of these figures, \emph{Wholesale} represents the cost for buying from the wholesale market,
\emph{Buying} represents the cost for buying on the hour-ahead market,
\emph{Selling} represents the revenue for selling on the hour-ahead market,
\emph{Penalty} represents the penalty for violating the selling contract,
and \emph{Overall} represents the overall reward of the sum of the costs and revenue.

As illustrated in Fig. \ref{fig:cost_type_mg_1}-\ref{fig:cost_type_mg_4}, the four microgrids can obtain a higher overall reward under our proposed method compared with the baselines.
The results show that the multiagent deep reinforcement learning approach can effectively improve the reward of each agent by collaboratively learning the policy for each agent compared with the baselines.

\begin{figure}
\begin{center}
\epsfig{file=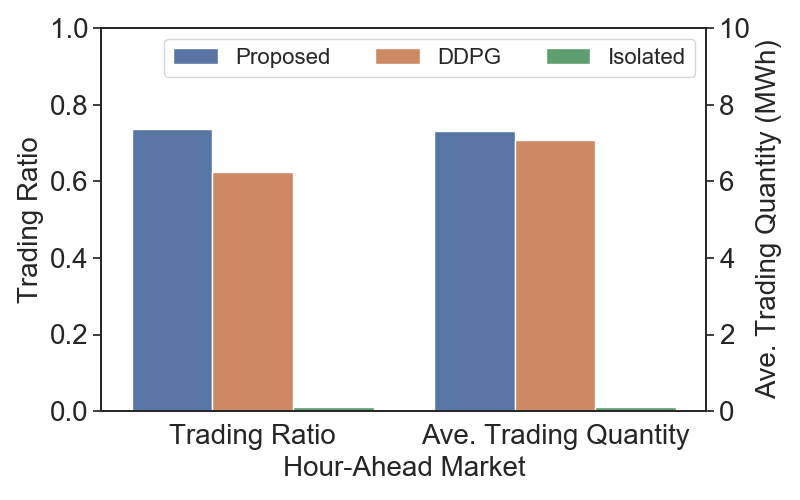, width=0.7\columnwidth}
\end{center}
\caption{The comparisons of the successful trading ratio and the average trading quantity per trading under different methods.} 
\label{fig:trading_cmp}
\end{figure}

As illustrated in Fig. \ref{fig:cost_type_mg_1} and \ref{fig:cost_type_mg_2},   
Microgrid 1 and 2 can obtain revenues by energy selling on the hour-ahead market.
Thus, the overall rewards under our method and DDPG are higher than Isolated.
With Isolated, the cost, revenue, and the overall reward are all near to zero, because Microgrid 1 and 2 have enough renewable energy generation to meet its demands, but cannot gain revenues without energy trading.
With our method, Microgrid 1 and 2 can gain more revenue compared with DDPG.
As illustrated in Fig. \ref{fig:trading_cmp}, 
this is partly due to that the successful trading ratio and the average trading quantity per time slot with our proposed method are higher than that of DDPG.
Therefore, Microgrid 1 and 2 can obtain more revenues on the hour-ahead market,
and Microgrid 3 and 4 can save more cost via energy trading.

Fig. \ref{fig:cost_type_mg_3} and \ref{fig:cost_type_mg_4} verify that our proposed method can reduce more cost for Microgrid 3 and 4 compared with the baseline methods.
This is because Microgrid 3 and 4 can successfully buy more energy from the hour-ahead market compared with DDPG, 
and the price on the hour-ahead market is much lower than the wholesale market,
thus it can save more cost compared with DDPG.
On the contrary, with the Isolated method, all energy must be brought from the wholesale market with a higher price,
it will incur more cost.

\section{Conclusion} \label{sec:conclusion}
In this paper, we propose a multiagent deep reinforcement learning approach for learning the optimal policy for distributed energy trading and scheduling among the microgrids.
Each microgrid is model as an agent, which can cooperate and compete with other agents on the hour-ahead energy market for energy trading and make the local scheduling decision.
We model the problem as a Markov game which aim to maximize the reward for each microgrid.
We adopt MADDPG to design the algorithm to learn the optimal policy for each agent.
To evaluate the performance, we conduct the experiments using the real-world datasets.
The experimental results show that each agent can find the equilibrium in the mixed cooperative-competitive environment and converge to the optimal policy.
Our method can reduce the cost and improve the overall reward for each microgrid compared with the baseline methods.

\ifCLASSOPTIONcaptionsoff
  \newpage
\fi



%

\bibliographystyle{IEEEtran}
\bibliography{IEEEsmartgrid}

\begin{thebibliography}{10}
\providecommand{\url}[1]{#1}
\csname url@samestyle\endcsname
\providecommand{\newblock}{\relax}
\providecommand{\bibinfo}[2]{#2}
\providecommand{\BIBentrySTDinterwordspacing}{\spaceskip=0pt\relax}
\providecommand{\BIBentryALTinterwordstretchfactor}{4}
\providecommand{\BIBentryALTinterwordspacing}{\spaceskip=\fontdimen2\font plus
\BIBentryALTinterwordstretchfactor\fontdimen3\font minus
  \fontdimen4\font\relax}
\providecommand{\BIBforeignlanguage}[2]{{%
\expandafter\ifx\csname l@#1\endcsname\relax
\typeout{** WARNING: IEEEtran.bst: No hyphenation pattern has been}%
\typeout{** loaded for the language `#1'. Using the pattern for}%
\typeout{** the default language instead.}%
\else
\language=\csname l@#1\endcsname
\fi
#2}}
\providecommand{\BIBdecl}{\relax}
\BIBdecl

\bibitem{lasseter2011smart}
R.~H. Lasseter, ``Smart distribution: Coupled microgrids,'' \emph{Proceedings
  of the IEEE}, vol.~99, no.~6, pp. 1074--1082, 2011.

\bibitem{mengelkamp2018designing}
E.~Mengelkamp, J.~G{\"a}rttner, K.~Rock, S.~Kessler, L.~Orsini, and
  C.~Weinhardt, ``Designing microgrid energy markets - a case study: {The}
  {Brooklyn} microgrid,'' \emph{Applied Energy}, vol. 210, pp. 870--880, 2018.

\bibitem{gregoratti2014distributed}
D.~Gregoratti and J.~Matamoros, ``Distributed energy trading: {The}
  multiple-microgrid case,'' \emph{iEEE Transactions on industrial
  Electronics}, vol.~62, no.~4, pp. 2551--2559, 2014.

\bibitem{matamoros2012microgrids}
J.~Matamoros, D.~Gregoratti, and M.~Dohler, ``Microgrids energy trading in
  islanding mode,'' in \emph{2012 IEEE Third International Conference on Smart
  Grid Communications (SmartGridComm)}.\hskip 1em plus 0.5em minus 0.4em\relax
  IEEE, 2012, pp. 49--54.

\bibitem{wang2014game}
Y.~Wang, W.~Saad, Z.~Han, H.~V. Poor, and T.~Ba{\c{s}}ar, ``A game-theoretic
  approach to energy trading in the smart grid,'' \emph{IEEE Transactions on
  Smart Grid}, vol.~5, no.~3, pp. 1439--1450, 2014.

\bibitem{lee2015distributed}
J.~Lee, J.~Guo, J.~K. Choi, and M.~Zukerman, ``Distributed energy trading in
  microgrids: A game-theoretic model and its equilibrium analysis,'' \emph{IEEE
  Transactions on Industrial Electronics}, vol.~62, no.~6, pp. 3524--3533,
  2015.

\bibitem{wang2016reinforcement}
H.~Wang, T.~Huang, X.~Liao, H.~Abu-Rub, and G.~Chen, ``Reinforcement learning
  in energy trading game among smart microgrids,'' \emph{IEEE Transactions on
  Industrial Electronics}, vol.~63, no.~8, pp. 5109--5119, 2016.

\bibitem{fathi2013adaptive}
M.~Fathi and H.~Bevrani, ``Adaptive energy consumption scheduling for connected
  microgrids under demand uncertainty,'' \emph{IEEE Transactions on Power
  Delivery}, vol.~28, no.~3, pp. 1576--1583, 2013.

\bibitem{huang2014adaptive}
Y.~Huang, S.~Mao, and R.~M. Nelms, ``Adaptive electricity scheduling in
  microgrids,'' \emph{IEEE Transactions on Smart Grid}, vol.~5, no.~1, pp.
  270--281, 2014.

\bibitem{fathi2013statistical}
M.~Fathi and H.~Bevrani, ``Statistical cooperative power dispatching in
  interconnected microgrids,'' \emph{IEEE Transactions on Sustainable Energy},
  vol.~4, no.~3, pp. 586--593, 2013.

\bibitem{wang2015bargaining}
H.~Wang and J.~Huang, ``Bargaining-based energy trading market for
  interconnected microgrids,'' in \emph{2015 IEEE International Conference on
  Communications (ICC)}.\hskip 1em plus 0.5em minus 0.4em\relax IEEE, 2015, pp.
  776--781.

\bibitem{paudel2018peer}
A.~Paudel, K.~Chaudhari, C.~Long, and H.~B. Gooi, ``{Peer-to-Peer} energy
  trading in a prosumer-based community microgrid: {A} game-theoretic model,''
  \emph{IEEE Transactions on Industrial Electronics}, vol.~66, no.~8, pp.
  6087--6097, 2018.

\bibitem{kim2019direct}
H.~Kim, J.~Lee, S.~Bahrami, and V.~Wong, ``Direct energy trading of microgrids
  in distribution energy market,'' \emph{IEEE Transactions on Power Systems},
  2019.

\bibitem{li2018distributed}
J.~Li, C.~Zhang, Z.~Xu, J.~Wang, J.~Zhao, and Y.-J.~A. Zhang, ``Distributed
  transactive energy trading framework in distribution networks,'' \emph{IEEE
  Transactions on Power Systems}, vol.~33, no.~6, pp. 7215--7227, 2018.

\bibitem{tushar2020peer}
W.~Tushar, T.~K. Saha, C.~Yuen, D.~Smith, and H.~V. Poor, ``Peer-to-peer
  trading in electricity networks: an overview,'' \emph{IEEE Transactions on
  Smart Grid}, 2020.

\bibitem{sutton2018reinforcement}
R.~S. Sutton and A.~G. Barto, \emph{Reinforcement learning: An
  introduction}.\hskip 1em plus 0.5em minus 0.4em\relax MIT press, 2018.

\bibitem{lowe2017multi}
R.~Lowe, Y.~Wu, A.~Tamar, J.~Harb, O.~P. Abbeel, and I.~Mordatch, ``Multi-agent
  actor-critic for mixed cooperative-competitive environments,'' in
  \emph{Advances in Neural Information Processing Systems}, 2017, pp.
  6379--6390.

\bibitem{wang2016cooperative}
H.~Wang and J.~Huang, ``Cooperative planning of renewable generations for
  interconnected microgrids,'' \emph{IEEE Transactions on Smart Grid}, vol.~7,
  no.~5, pp. 2486--2496, 2016.

\bibitem{weng2018distributed}
S.~Weng, D.~Yue, C.~Dou, J.~Shi, and C.~Huang, ``Distributed event-triggered
  cooperative control for frequency and voltage stability and power sharing in
  isolated inverter-based microgrid,'' \emph{IEEE transactions on cybernetics},
  no.~99, pp. 1--13, 2018.

\bibitem{ahn2017distributed}
H.-S. Ahn, B.-Y. Kim, Y.-H. Lim, B.-H. Lee, and K.-K. Oh, ``Distributed
  coordination for optimal energy generation and distribution in cyber-physical
  energy networks,'' \emph{IEEE transactions on cybernetics}, vol.~48, no.~3,
  pp. 941--954, 2017.

\bibitem{wang2016incentivizing}
H.~Wang and J.~Huang, ``Incentivizing energy trading for interconnected
  microgrids,'' \emph{IEEE Transactions on Smart Grid}, vol.~9, no.~4, pp.
  2647--2657, 2016.

\bibitem{littman1994markov}
M.~L. Littman, ``Markov games as a framework for multi-agent reinforcement
  learning,'' in \emph{Machine learning proceedings 1994}.\hskip 1em plus 0.5em
  minus 0.4em\relax Elsevier, 1994, pp. 157--163.

\bibitem{uhlenbeck1930theory}
G.~E. Uhlenbeck and L.~S. Ornstein, ``On the theory of the brownian motion,''
  \emph{Physical review}, vol.~36, no.~5, p. 823, 1930.

\bibitem{solcast}
``{Solcast},'' \url{https://toolkit.solcast.com.au/}, accessed: May 2020.

\bibitem{load_sg}
``{Half Hourly System Demand},''
  \url{https://data.gov.sg/dataset/half-hourly-system-demand}, accessed: May
  2020.

\end{thebibliography}

\end{document}